\documentclass[a4paper]{jpconf}
\usepackage{graphicx}
\usepackage{amsmath}
\usepackage{amsfonts}
\usepackage{amssymb}
\usepackage{txfonts}
\usepackage{bm}
\begin{document}

\title{Evidence for the Sr$_2$RuO$_4$ intercalations 
in the Sr$_3$Ru$_2$O$_7$ region of the Sr$_3$Ru$_2$O$_7$\,-\,Sr$_2$RuO$_4$ eutectic system}

\author{S.~Kittaka$^a$, S.~Yonezawa$^a$, H.~Yaguchi$^{a,b}$, Y.~Maeno$^a$, R.~Fittipaldi$^{a,c}$, A.~Vecchione$^{a,c}$, 
J.~-F.~Mercure$^d$, A.~Gibbs$^d$, R.~S.~Perry$^{d,e}$, and A.~P.~Mackenzie$^d$}

\address{$^a$Department of Physics, Graduate School of Science, Kyoto University, Kyoto 606-8502, Japan}
\address{$^b$Department of Physics, Faculty of Science and Technology, Tokyo University of Science, Noda 278-8510, Japan}
\address{$^c$CNR-INFM Regional Laboratory ``SuperMat'' and Department of Physics, University of Salerno, I-84081 Baronissi (Sa), Italy}
\address{$^d$School of Physics and Astronomy, University of St. Andrews, St. Andrews KY16 9SS, UK}
\address{$^e$School of Physics, University of Edinburgh, Edinburgh EH9 3JZ, UK}

\ead{kittaka@scphys.kyoto-u.ac.jp}

\begin{abstract}
Although Sr$_3$Ru$_2$O$_7$ has not been reported to exhibit superconductivity so far, 
ac susceptibility measurements revealed multiple superconducting transitions occurring in the Sr$_3$Ru$_2$O$_7$ region cut from Sr$_3$Ru$_2$O$_7$\,-\,Sr$_2$RuO$_4$ eutectic crystals. 
Based on various experimental results, some of us proposed the scenario 
in which Sr$_2$RuO$_4$ thin slabs with a few layers of the RuO$_2$ plane are embedded in the Sr$_3$Ru$_2$O$_7$ region as stacking faults and 
multiple superconducting transitions arise from the distribution of the slab thickness.
To examine this scenario, 
we measured the resistivity along the $ab$ plane ($\rho_{ab}$) using a Sr$_3$Ru$_2$O$_7$-region sample cut from the eutectic crystal, as well as along the $c$ axis ($\rho_{c}$) using the same crystal.
As a result, we detected resistance drops associated with superconductivity only in $\rho_{ab}$, but not in $\rho_{c}$. 
These results support the Sr$_2$RuO$_4$ thin-slab scenario.
In addition, we measured the resistivity of a single crystal of pure Sr$_3$Ru$_2$O$_7$ with very high quality and 
found that pure Sr$_3$Ru$_2$O$_7$ does not exhibit superconductivity down to 15~mK.

\end{abstract}

\section{Introduction}
The Ruddlesden-Popper series of layered perovskites Sr$_{n+1}$Ru$_n$O$_{3n+1}$ is a fascinating research subject because 
the $n$=1 member Sr$_2$RuO$_4$ is now believed to be a spin-triplet superconductor \cite{Maeno1994Nature,Mackenzie2003RMP}. 
The $n$=2 member of this series Sr$_3$Ru$_2$O$_7$ is known to exhibit an enhanced Pauli paramagnetism with a metamagnetic transition \cite{Ikeda2000PRB,Perry2001PRL,Borzi2004PRL}. 
Although crystals of Sr$_3$Ru$_2$O$_7$ are highly refined, the superconductivity in Sr$_3$Ru$_2$O$_7$ has not been discovered so far. 
Recently, Sr$_3$Ru$_2$O$_7$-Sr$_2$RuO$_4$ eutectic crystals were successfully grown \cite{Fittipaldi2005JCG} and 
surprisingly, multiple superconducting transitions were observed in the ac susceptibility measurements using a Sr$_3$Ru$_2$O$_7$-region sample cut from eutectic crystals \cite{Kittaka2008PRB}, as presented in Fig.~\ref{ec}(a).
However, it has been revealed that this superconductivity is not a bulk property of Sr$_3$Ru$_2$O$_7$
because these superconducting transitions are easily suppressed by small ac magnetic fields and no anomaly was observed in the specific heat \cite{Kittaka2008PRB}. 
The most plausible scenario of this superconductivity is that  
Sr$_2$RuO$_4$ thin slabs with a few layers of RuO$_2$ planes are embedded in the Sr$_3$Ru$_2$O$_7$ region and the multiple superconducting transitions arise from the distribution of the slab thickness.

In order to obtain additional evidence for the origin of the superconductivity observed in the Sr$_3$Ru$_2$O$_7$ region cut from eutectic crystals, 
we measured the resistivity along the $ab$ plane ($\rho_{ab}$) of a Sr$_3$Ru$_2$O$_7$-region sample cut from the eutectic crystal, 
which we designate below as a eutectic Sr$_3$Ru$_2$O$_7$ sample, 
as well as along the $c$ axis ($\rho_{c}$) of the same sample.
In addition, we measured $\rho_{c}$ of a pure (i.e. non-eutectic) Sr$_3$Ru$_2$O$_7$ sample with very high quality 
in order to clarify whether or not pure Sr$_3$Ru$_2$O$_7$ exhibits superconductivity at low temperatures.

\section{Experimental}

Resistivity was measured using a conventional four-probe method with an ac current.
Single crystals of the Sr$_3$Ru$_2$O$_7$-Sr$_2$RuO$_4$ eutectic system and those of single-phase Sr$_3$Ru$_2$O$_7$ were grown by a floating-zone method.
The size of the eutectic Sr$_3$Ru$_2$O$_7$ sample was approximately 1.5 $\times$ 0.7 mm$^2$ in the $ab$ plane and 0.3 mm along the $c$ axis. 
The results of ac susceptibility and specific heat measurements using this eutectic Sr$_3$Ru$_2$O$_7$ sample were reported in Ref.~\cite{Kittaka2008PRB}.
The dimensions of the pure Sr$_3$Ru$_2$O$_7$ sample are 0.44 $\times$ 0.28 mm$^2$ in the $ab$ plane and 1.14 mm along the $c$ axis.
The $\rho_{ab}$ measurements on the eutectic Sr$_3$Ru$_2$O$_7$ sample were performed down to 0.3~K with a $^3$He cryostat (Oxford Instruments, model Heliox VL). 
After the $\rho_{ab}$ measurements, we removed the electrical leads and attached another set of wires again on the same sample and performed $\rho_c$ measurements down to 0.1~K with an adiabatic demagnetization refrigerator (Cambridge Magnetic Refrigeration, mFridge50).
The $\rho_c$ measurements using the pure Sr$_3$Ru$_2$O$_7$ sample were performed with a $^3$He-$^4$He dilution refrigerator (Cryoconcept, model DR-JT-S-100-10) down to 15~mK. 
In this study, we used a cylinder of permalloy (Hamamatsu Photonics K.K., E989-28) in order to reduce remanent fields such as the earth field.

\section{Results and Discussion}
\subsection{Resistivity of the eutectic Sr$_3$Ru$_2$O$_7$}

\begin{figure}[h]
\begin{center}
\includegraphics[width=5.8in]{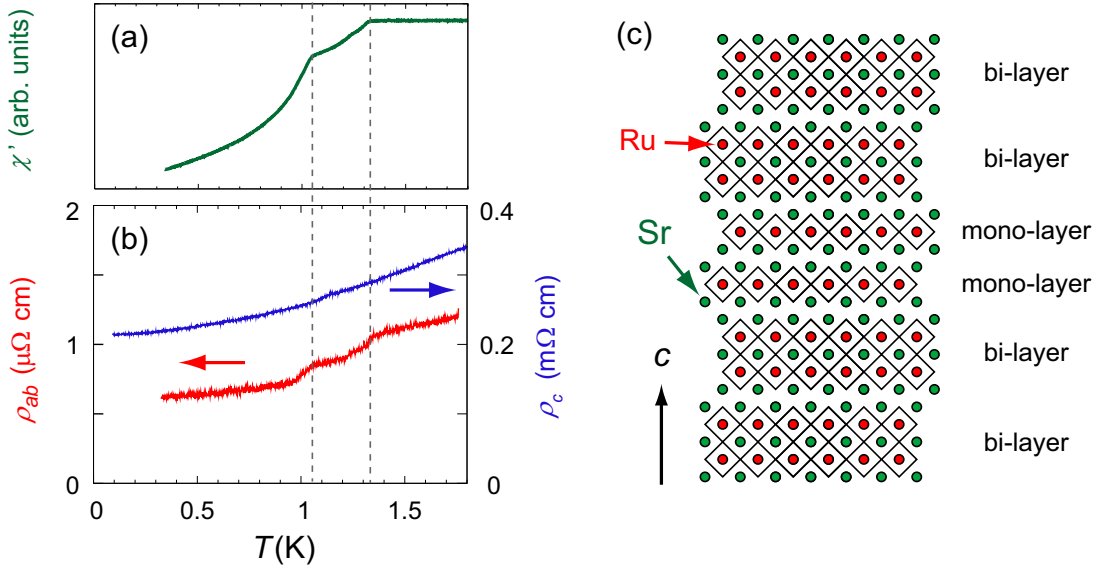}\hspace{2pc}
\caption{\label{ec}(a) Temperature dependence of the real part of the ac susceptibility for a eutectic Sr$_3$Ru$_2$O$_7$ sample with $\mu_0H_\mathrm{ac}$=0.58 $\muup$T and $f$=3011 Hz(sample~1 in Ref.~\cite{Kittaka2008PRB}). 
(b) Temperature dependences of the resistivity along the $c$ axis and that along the $ab$ plane for the eutectic Sr$_3$Ru$_2$O$_7$ sample measured at $f$=89.1 Hz ($I$=0.5 mA-rms for $\rho_{ab}$ and $I$=0.1 mA-rms for $\rho_{c}$).
(c) A schematic image of two monolayers of RuO$_2$ planes (Sr$_2$RuO$_4$) intercalated in bilayers (Sr$_3$Ru$_2$O$_7$). 
Oxygens are located at the corner of the octahedra.}
\end{center}
\end{figure}

Figure \ref{ec}(b) shows temperature dependence of $\rho_{ab}$ and $\rho_c$ of the eutectic Sr$_3$Ru$_2$O$_7$ sample.
The values of $\rho_{ab}$ and $\rho_c$ are approximately 1~$\muup \Omega$cm and 300 $\muup \Omega$cm at 1.5~K, respectively.
The in-plane resistivity is nearly the same as those of pure Sr$_3$Ru$_2$O$_7$ crystals with very high quality \cite{Borzi2004PRL}.
This low value of the in-plane resistivity indicates that Sr$_3$Ru$_2$O$_7$ in the eutectic system crystallized with high quality.
Also, macroscopic Sr$_2$RuO$_4$ domains in the eutectic system are high quality 
because its $T_\mathrm{c}$ is nearly 1.5~K \cite{Kittaka2008PRB},
which is one of the best $T_\mathrm{c}$ of Sr$_2$RuO$_4$ reported.
It is interesting that both Sr$_2$RuO$_4$ and Sr$_3$Ru$_2$O$_7$ spontaneously crystallize with high quality in this eutectic system.

In $\rho_{ab}$ measurements, two clear resistance drops were observed at 1.05 and 1.32~K. 
These transition temperatures well coincide with those observed in the ac susceptibility [Fig.~\ref{ec}(a)].
However, in $\rho_c$ measurements, no obvious transition was observed.
These results are consistent with the Sr$_2$RuO$_4$ thin-slab scenario because
they imply that superconducting inclusions embedded in the eutectic Sr$_3$Ru$_2$O$_7$ are too thin along the $c$ axis to shortcircuit the current path along the interlayer direction.
This behavior is in sharp contrast with $\rho_c$ for the Sr$_2$RuO$_4$-Ru eutectic system, 
in which the emergence of the ``3-K" superconductivity in the interface of Ru lamellae results in a large drop in $\rho_c$ \cite{Maeno1998PRL}.
Although we observed two clear transitions in the $\rho_{ab}$ measurement using the eutectic Sr$_3$Ru$_2$O$_7$ sample, 
$\rho_{ab}$ does not become zero down to low temperatures.
This non-zero resistivity implies that 
Sr$_2$RuO$_4$ inclusions in this eutectic Sr$_3$Ru$_2$O$_7$ sample do not completely form a path between the voltage contacts.
The Sr$_2$RuO$_4$ inclusions are probably well separated in this sample.
In some cases, eutectic Sr$_3$Ru$_2$O$_7$ samples exhibit zero resistivity (e.~g., Ref. \cite{Fittipaldi2008EPL}),
probably because Sr$_2$RuO$_4$ inclusions in such samples link a path between the voltage contacts. 
Now, on the basis of various experiments, we believe that the origin of the superconductivity observed in the eutectic Sr$_3$Ru$_2$O$_7$ sample is 
the presence of several monolayers of RuO$_2$ planes intercalated in Sr$_3$Ru$_2$O$_7$ as stacking faults. 
For example, two monolayers of RuO$_2$ planes intercalated in Sr$_3$Ru$_2$O$_7$ is schematically drawn in Fig.~\ref{ec}(c).
In fact, such stacked monolayers of RuO$_2$ planes have been observed with a transmission electron microscope \cite{Fittipaldi2008EPL}.

\subsection{Resistivity of pure Sr$_3$Ru$_2$O$_7$ with high quality}
We are also interested in the possibility of superconductivity in pure Sr$_3$Ru$_2$O$_7$.
In order to examine the superconductivity in pure Sr$_3$Ru$_2$O$_7$, we consider it important
(i) to use single crystals of single-phase Sr$_3$Ru$_2$O$_7$ with very high quality, 
(ii) to cool down the sample to sufficiently low temperature, and 
(iii) to perform measurements in zero field by excluding the geomagnetic field.
Therefore, we measured $\rho_c$ down to 15~mK using a single crystal of pure Sr$_3$Ru$_2$O$_7$ with the in-plane residual resistivity of 0.4 $\muup \Omega$cm, 
which is one of the highest-quality Sr$_3$Ru$_2$O$_7$ grown so far. 
In addition, by placing the sample in a cylinder of permalloy, we reduced the residual field to be lower than 0.1~$\muup$T.

\begin{figure}[h]
\begin{center}
\includegraphics[width=3.5in]{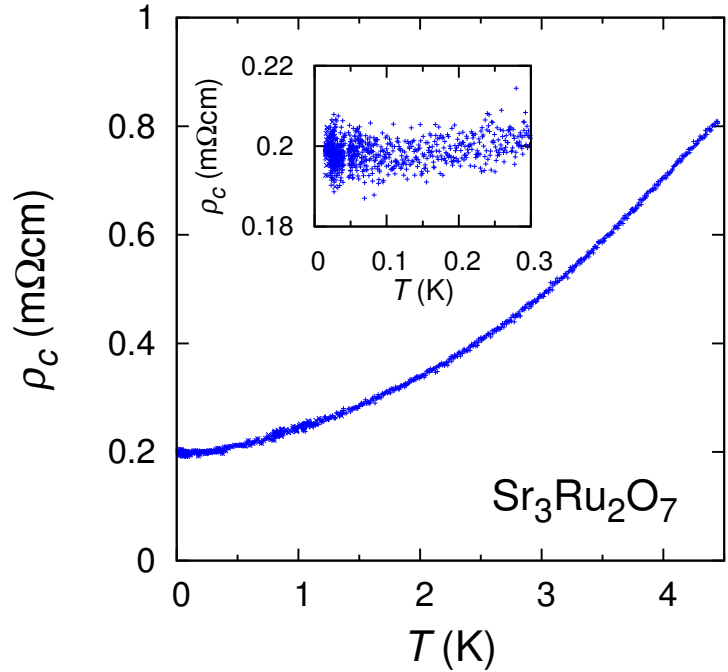}
\hspace{5pc}
\caption{\label{sp}Temperature dependence of the resistivity along the $c$ axis of 
the pure Sr$_3$Ru$_2$O$_7$ sample with very high quality measured at $I$=0.01 mA-rms with $f$=7 Hz without the geomagnetic field.
The inset shows the low-temperature region below 0.3~K.}
\end{center}
\end{figure}

Figure~\ref{sp} shows temperature dependence of $\rho_c$ of the pure Sr$_3$Ru$_2$O$_7$ sample.
The $\rho_c$ monotonically decreases with decreasing temperature and 
no anomaly indicating a superconducting transition was observed down to 15~mK.
From this measurement, we conclude that pure Sr$_3$Ru$_2$O$_7$ does not become superconducting down to 15~mK.

\section{Conclusion}
We measured the resistivity along the $c$ axis as well as along the $ab$ plane of a eutectic Sr$_3$Ru$_2$O$_7$ sample.
The resistance drops due to the multiple superconducting transitions were observed only for $\rho_{ab}$, but not for $\rho_c$.
This result indicates that superconductors with thin thickness along the $c$ axis are embedded in the eutectic Sr$_3$Ru$_2$O$_7$. 
Now, we are convinced that the origin of the superconductivity observed in the eutectic Sr$_3$Ru$_2$O$_7$ sample is the presence of Sr$_2$RuO$_4$ inclusions embedded in Sr$_3$Ru$_2$O$_7$ as stacking faults.
In order to search for superconductivity in pure Sr$_3$Ru$_2$O$_7$, 
we also measured the resistivity along the $c$ axis using a single crystal of best-quality pure Sr$_3$Ru$_2$O$_7$.
However, no resistance anomaly associated with the superconducting transition was observed down to 15~mK.
Therefore, we conclude that pure Sr$_3$Ru$_2$O$_7$ does not become superconducting down to 15~mK.

\section*{Acknowledgement}
This work has been supported by the Grant-in-Aid for the Global COE program ``The Next Generation of Physics, Spun from Universality and Emergence'' 
from the Ministry of Education, Culture, Sports, Science, and Technology (MEXT) of Japan.
It is also supported by Grants-in-Aid for Scientific Research from MEXT and from the Japan Society for the Promotion of Science (JSPS).
One of the authors (S. K.) is financially supported as a JSPS Research Fellow.

\section*{References}
\providecommand{\newblock}{}

\end{document}